\documentstyle[12pt]{article} 
\begin{document}
\title{One-dimensional quantum chaos:
       Explicitly solvable cases}
\author{Yu. Dabaghian, R. V. Jensen and R. Bl\"umel \cr
    Department of Physics, Wesleyan University, \cr
     Middletown, CT 06459-0155, USA, \cr
     ydabaghian@wesleyan.edu}

\date{\today}
\maketitle

\begin{abstract}
We present quantum graphs with remarkably
regular spectral characteristics. We call them {\it regular quantum graphs}.
Although regular quantum graphs are strongly chaotic in the classical limit,
their quantum spectra are explicitly solvable in terms of
periodic orbits. We present
analytical solutions for the spectrum of regular quantum
graphs in the form of explicit and exact periodic orbit
expansions for each individual energy level.
\end{abstract}


\section{Introduction}

Consider a point particle moving along a network of bonds and vertices.
Schematically, the network is represented by a graph $\Gamma$
(see Fig.~1 for an example), which consists of $N_{B}$ bonds and $N_{V}$ 
vertices. 
The vertices are denoted by $V_i$; a bond connecting
vertices $i$ and $j$ is denoted by $B_{ij}$. The set of bonds
and vertices of $\Gamma$ defines its {\it geometry}.
We define a set of the bond potentials, $U_{ij}(k,x)$, where $x$ and
$k$ are correspondingly the coordinate and the momentum of
the particle on the bond $B_{ij}$.
The vertices of $\Gamma$ may be equipped with $\delta$-sources, etc.
The geometry of $\Gamma$ does not uniquely define the
dynamics of a particle on $\Gamma$. In fact, since for any given
geometry the graph may be ``dressed'' with arbitrary
bond and vertex potentials, there exist infinitely many
``dynamical realizations'' of $\Gamma$. We call
the set of bond and vertex potentials
the ``dynamical dressing'' of the graph.
Previously \cite{Roth,Orsay,QGT1,QGT2,QGT3}, mainly the ``bare-bond''
graphs were studied, where the particle moves freely on the bonds.

In this paper we focus on cases that have
no turning points on the bonds, i.e.
the energy of
the particle is larger than all of the bond potentials, $E>U_{ij}(x,k)$,
$x\in B_{ij}$.
A simple way to implement this condition is to
require that
the system is {\em scaling}
\cite{RS2,Koch,Bauch,Found,Nova}. This implies
$U_{ij}(x,k)=\lambda_{ij}(x)\, k^{2}$,
where the functions $\lambda_{ij}(x)$ are bounded for all $x$.
In this paper 
we consider only simple cases where the functions $\lambda_{ij}(x)$
are $x$ independent constants,
\begin{equation}
U_{ij}(x,k)=\lambda_{ij} k^{2}.
\label{const}
\end{equation}
This is very similar to moving on a free graph except 
for substituting the bond lengths with the action lengths
\begin{equation}
S^{0}_{ij}=\beta_{ij}L_{ij},
\label{action}
\end{equation}
where $L_{ij}$ is the length of the bond $B_{ij}$, and
$\beta_{ij}=\sqrt{1-\lambda_{ij}}$.
The scaling assumption (\ref{const}) is not an oversimplification of
the problem. Plenty of room is left
for very interesting phenomena. Moreover,
scaling quantum systems of this kind
are the analogues of certain electromagnetic ray-splitting systems
which have already been investigated experimentally in the
laboratory
\cite{Koch,Bauch,Found}.

For all but the most trivial graphs, i.e. linear or circular graphs with
vanishing bond and vertex potentials, the classical motion
on a graph, independently of
any particular dressing, is fully chaotic with positive topological entropy
\cite{Gutzw}. This means that
the number of possible periodic orbits traced by the particle
increases exponentially with their lengths. If no dynamical
turning points are present, the topological entropy
is independent of the
dynamical dressing and depends only on the geometry of the graph.
Since at any vertex different from a ``dead-end'' vertex
the classical particle has to choose randomly between several
possibilities (reflection, transmission, branching),
the particle's dynamical evolution resembles a stochastic, Markovian
process.

Given their classical chaoticity
it is surprising that the density of states
of quantum graphs
can be obtained exactly in terms of periodic orbit expansion series
\cite{Roth,QGT1,QGT2,QGT3}. 
Furthermore,
quantum graphs are considerably ``more
integrable'' than all the previously known exactly solvable
quantum systems. For example we will show below
that for a certain class of quantum graphs --- we call them
{\it regular quantum graphs} --- there exists an explicit and exact
periodic orbit expansion for every quantum energy level. 
In other words,
although the classical limit of regular quantum graphs is chaotic,
each individual level of their spectra can be obtained
exactly and explicitly via an analytical formula containing an
explicit sum over the periodic orbits of the graph.
To the authors' knowledge this is the first time that the spectrum of
a quantum chaotic system is obtained both {\it exactly} and {\it explicitly}.

The formal definition of regular quantum graphs is based on the
properties of the spectral equation \cite{QGT1,QGT2,QGT3}
\begin{equation}
\det [1-S(k)]=0,
\label{det}
\end{equation}
where $S(k)$ is the scattering matrix of the
graph \cite{QGT1}.
The modulus of the complex function (\ref{det}) is a
trigonometric polynomial of the form
\begin{equation}
\cos(S_0 k-\pi\gamma_{0})-\Phi(k) = 0,
\label{eqn}
\end{equation}
where
\begin{equation}
\Phi(k)=\sum_{i}a_{i}\cos(S_{i}k-\pi\gamma_{i}).
\label{Phi}
\end{equation}
and 
\begin{equation}
S_{0}={1\over k}\sum_{i<j}\, \int_{B_{ij}}\, k_{ij}(x)\, dx
\label{S0}
\end{equation} 
is the total reduced action length of the graph $\Gamma$ and 
the constant frequencies $S_{i}<S_{0}$ naturally emerge as 
combinations of the reduced classical actions (\ref{action}). 
Under the scaling assumption, the coefficients $a_{i}$, $\gamma_{0}$ 
and $\gamma_{i}$ are constants. 

We now define regular quantum graphs. They satisfy
\begin{equation}
\alpha=\sum_i\mid a_i|<1.
\label{regul}
\end{equation}
The motivation for this definition is the following:
it allows us to solve (\ref{eqn}) formally for the momentum
eigenvalues $k_{n}$,
\begin{equation}
k_{n}={\pi\over S_0}\left[n+\mu+\gamma_{0}\right] - {1\over S_0}
\cases{\arccos(\Phi(k_n)), &for $n+\mu$ even \cr
        \pi-\arccos(\Phi(k_n)), &for $n+\mu$ odd, \cr}
\label{levels}
\end{equation}
where $\mu$ is a fixed integer chosen such that $k_1$ is
the first non-negative solution of (\ref{eqn}).
Because of
(\ref{regul}) the second term
of (\ref{levels}) assumes only values
between $u$ and $\pi/S_0-u$, where $0<u=\arccos(\alpha)/S_0<\pi/2S_0$.
Thus, for regular graphs, the points
\begin{equation}
\bar{k}_{n}=\frac{\pi}{S_0}(n+\gamma), \ \ n=1,2,\ldots ,
\ \ \ \gamma\equiv\gamma_{0}+\mu,
\label{sep}
\end{equation}
are guaranteed not to be roots of (\ref{det}) and serve as
separators between roots number $n$ and $n+1$. 

Obviously the function (\ref{sep}) reflects the average behavior of the 
levels of the momentum. It is simply the inverted average staircase,
$\bar{k}_{N}=\bar N(k)^{-1}$.
Geometrically the points (\ref{sep}) are the intersection points 
between the staircase function,
$N(k)\equiv\sum_{n}\Theta(k-k_{n})$, and the average
staircase $\bar N(\bar k)$
resulting in the crossing condition
\begin{equation}
\bar N(\bar k_n)=N(\bar k_n)=n.
\label{cross}
\end{equation}
The crossing condition (\ref{cross})
is illustrated in (Fig.~3).
In case if there are no source potentials defined at the graph vertices,
it can be shown that
\begin{equation}
\gamma=-\bar N(0)=\frac{1}{2}.
\label{gamma}
\end{equation}

The existence of the separating points (\ref{sep}) implies that the 
roots (\ref{levels}) are confined to the ``root zones'', or ``root 
intervals'' $I_n=[\bar k_{n-1},\bar k_n]$, $n=1,2,\ldots$.
If $\alpha\leq C<1$ holds ($C$ constant), equation (\ref{levels}) 
implies the existence of finite-width root-free ``forbidden zones''
$R_n=(\bar k_n-u, \bar k_n+u)$ surrounding every separating point 
$\bar k_n$, where no roots of (\ref{det}) can be found.
The roots of (\ref{det}) can only be found in the ``allowed zones''
$Z_n=[\bar k_{n-1}+u,\bar k_n-u]$, which are subsets of the root 
intervals $I_n=[\bar k_{n-1},\bar k_n]$. For $C\rightarrow 1$ the
width of the forbidden regions shrinks, $u\rightarrow 0$, and the 
allowed zones occupy the whole interval, $Z_n\rightarrow I_n$.

Since $S_0$ is the largest action in (\ref{eqn}) and (\ref{Phi}),
it can be shown \cite{Graf} that $k_n$ is the {\it only} root
in $Z_n$.
Therefore there is exactly one root $k_n$ inside of $Z_n\subset I_n$, 
and this root is bounded away from the separating points 
$\bar k_{n-1}$ and $\bar k_n$ by a finite interval of length $2u$.

The existence of the separating points (\ref{sep}) and the root-free
zones $R_n$ are the key for obtaining an explicit and exact periodic 
orbit expansion for every root of (\ref{det}).
The starting point for obtaining the explicit expressions
is the exact periodic orbit expansion for the density of states,
\begin{equation}
\rho (k)\equiv \sum_{j=1}^{\infty}\delta \left(k-k_{j}\right).
\label{ro}
\end{equation}
As shown in \cite{Roth,QGT1,QGT2,QGT3} it can be written explicitly as
\begin{equation}
\rho (k) = \bar{\rho}(k)+\frac{1}{\pi}\mathop{\rm Re}
\sum_{p}S_{p}^{0}\sum_{\nu=1}^{\infty}A_{p}^{\nu}\,e^{i\nu S_{p}^{0}k}.
\label{rho}
\end{equation}
Here $\bar{\rho}(k)$ is the average density of states,
$\nu$ is the
repetition index, and $S_{p}^{0}$, $A_{p}$ are
correspondingly the reduced
action and the weight factor of the prime
periodic orbit labeled by $p$.
In the scaling case, $S_{p}^{0}$ and $A_{p}$
are $k$-independent constants \cite{Graf}.
Multiplying the density of states by $k$ and integrating from
$\bar k_{n-1}$ to $\bar k_{n}$ yields the
value of the root contained between
these separating points,
\begin{equation}
\int_{\bar k_{n-1}}^{\bar k_{n}}\rho (k)k\, dk=
\int_{\bar k_{n-1}}^{\bar k_{n}}\sum_{j=1}^{\infty}
\delta\left(k-k_{j}\right)k\,dk
=k_{n}.
\label{root}
\end{equation}
Performing the same procedure using the series expansion
representation (\ref{rho}) and the crossing condition (\ref{cross}),
we obtain
\begin{eqnarray}
k_{n}=\frac{\pi}{S_{0}}n-\frac{2}{\pi}\sum_{p}
\frac{1}{S_{p}^{0}}\sum_{\nu=1}^{\infty}
\frac{A_{p}^{\nu}}{\nu^{2}}\sin(\frac{\pi}{2}\nu\omega_{p})\,
\sin(\pi\nu\omega_{p}n),
\label{kn}
\end{eqnarray}
where $\omega_{p}=S_{p}^{0}/S_0$, and the $A_{p}$'s are assumed to be real
(no vertex potentials).
 
Since all of the quantities on the right-hand side of (\ref{kn}) are 
known, this formula provides an explicit representation of the roots 
$k_n$ of the spectral equation (\ref{det}) in terms of the geometric 
and dynamical characteristics of the graph. 
To our knowledge, this is the first time that the energy levels of a 
chaotic system are expressed explicitly in terms of a periodic orbit
expansion. Previously, explicit formulae for individual energy levels 
were known only for integrable systems.
In the context of periodic orbit theory, the energy levels of 
integrable systems are given by the Einstein-Brillouin-Keller (EBK) 
formula \cite{Gutzw}.
However, apart from a few exceptional cases \cite{Szabo} EBK 
quantization is only of semiclassical accuracy.

The difference between (\ref{rho}) and (\ref{kn}) is profound.
The density of states (\ref{rho}) allows the computation of spectral 
points only indirectly as the singularities of (\ref{rho}).
Formula (\ref{kn}), on the other hand, allows the computation of every 
quantum level {\it individually}, {\it explicitly} and {\it exactly}
in terms of classical parameters.

In order to demonstrate that the class of regular quantum graphs is 
not empty we present an explicit example: the one-dimensional scaled 
step potential with $V_0=\lambda E$.
A sketch of this potential is shown in Fig.~3.
Physically this potential is realized, e.g., by a rectangular microwave 
cavity partially loaded with a dielectric substance \cite{Koch,Bauch,Found}.
The scaling step potential is equivalent to the scaling three-vertex 
linear graph shown in Fig. 2 (b). It has two bonds $L_{1}=b$ and
$L_{2}=\beta(1-b)$; the single scaling constant $\beta$ (see (\ref{action})) 
is given by $\beta=\sqrt{1-\lambda}$. The spectral equation is given by
\begin{equation}
\left|\det\left[ 1-S(k)\right]\right|=
\sin(Lk)-r\sin\left[\left(L_{1}-L_{2}\right)k\right] =0,
\label{3hydra}
\end{equation}
where $L=L_{1}+L_{2}$, and $r=(1-\beta)/(1+\beta)$
is the reflection coefficient at the vertex $V_{2}$ between the two bonds.
It defines the eigenvalues $k_n$ only implicitly and is usually solved by 
graphical or numerical methods.
Application of (\ref{kn}), however, solves (\ref{3hydra}) explicitly in 
terms of periodic orbits such as the ones shown in (Fig.~2). In order to 
apply (\ref{3hydra}) we need the coefficients $A_{p}$. 
They are given by \cite{Nova,Graf}
\begin{equation}
A_{p}= (-1)^{\chi(p)}r^{\sigma(p)}(1-r^{2})^{\tau(p)/2},
\label{A}
\end{equation}
where $r$ is the reflection coefficient at the middle vertex and $\sigma(p)$
and $\tau(p)$ are correspondingly the number of the reflections and
the transmissions through it.
Since the reflection coefficient may be positive or negative depending on
whether the particle scatters from the right or from the left, the factor
$(-1)^{\chi(p)}$ is needed to keep track of how many times it appears with 
a minus sign, including the sign changes due to the wall ($x=0$ and $x=1$)
reflections.

In order to illustrate the convergence of (\ref{kn}) we computed $k_1$,
$k_{10}$ and $k_{100}$ of the scaling step potential including periodic 
orbits of increasing binary length $q$. 
For the parameters of the potential we chose $b=0.3$ (see Fig. 2) and 
$\lambda=1/2$. Figure~4 shows the relative error 
$\epsilon_n^{(q)}=|k_n^{(q)}-k_n^{\rm exact}|/k_n^{\rm exact}$
for $n=1,10,100$ and $q$ ranging from 1 to 150.
We see that even for small $q$ the relative error is very small, decreasing 
further for large $q$ as a power-law in $q$. The power of convergence appears 
to be the same for all three $k$ and is close to $-2$. 
The convergence with $q$ is an important result. It indicates that although
(\ref{kn}) is only conditionally convergent it (i) converges to the correct 
result and (ii) is not just asymptotically convergent, but keeps converging 
when more and more periodic orbits are included.

Additional examples of regular quantum graphs are provided by all 
linear and circular quantum graphs with at most two bonds per vertex, 
independently of the number of vertices.
In other words, for any simply connected quantum graph and any dynamical
dressing there always exists a set of scaling constants $\lambda_{ij}$ of 
finite measure such that the regularity condition (\ref{regul}) is fulfilled.
Well-known particular cases of these simply connected quantum graphs are the 
``Manhattan potentials'', which are obvious generalizations of the simple step 
potential shown in Fig.~2 (a) to arbitrarily many steps inside of the well, and
linear chain graphs with scaling $\delta$ function potentials at the vertices.

It should be emphasized that the ``inverse staircase expansion'' (\ref{kn}) 
is not just a curious finding, valid for some simple $1D$ systems such as 
quantum graphs.
Similar explicit series may be obtained for more complicated higher dimensional 
systems when the following two key ingredients are available. 
The first ingredient is the exact series expansion of the density of states
(\ref{rho}), which has already been established for other classically chaotic
systems such as, e.g., quantum billiards \cite{AM}.
The second ingredient is a (piercing) average staircase function
$\bar N(k)$ or the inverted staircase function $\bar k_{n}$ {\em which intersects 
every stair of the staircase}, $\bar N(\bar k_{n})= N(\bar k_{n})=n$, $n=1,2,...$.
The intersection points $k_{n}$ then serve as the separators for the
possible root locations, and the procedure outlined in the text can be used
to find the periodic orbit expansions for individual roots of the system at hand.
In most cases, of course,
it is highly nontrivial to obtain 
these two necessary ingredients. 
The quantum graphs themselves are an excellent
illustration of this point.
While the expansion (\ref{rho}) is valid
for all quantum graphs, it is
the crossing condition (\ref{cross}) that is violated
when the inequality (\ref{regul}) brakes down.
The regular graphs are precisely those
for which the line $\bar N(k)=S_0k/\pi+\bar N(0)$
satisfies (\ref{cross})
and allows the application of the analytical
procedure that resulted in the explicit
formula (\ref{kn}) for the representation and
computation of
individual eigenvalues $k_n$.

Y.D. and R.B. gratefully acknowledge financial
support by NSF grants No.
PHY-9900730 and PHY-9984075; Y.D. and R.J by NSF
grant No. PHY-9900746.

%%%%%%%%%%%%%%%%%%%%%%%%%%%%%%%%%%%%%%%%%%%%%%%%%%%%%%%%%%%%%%%%

\pagebreak

\centerline{\bf Figure Captions}

\bigskip \noindent
{\bf Fig.~1:} A generic (quantum) graph with five vertices and
  six bonds.

\bigskip \noindent
{\bf Fig.~2:} (a) Simple step potential, a basic problem in
  one-dimensional quantum mechanics.
  Also shown are examples of Newtonian (``N'') and
  non-Newtonian (``NN'') periodic orbits used in
  the periodic orbit expansion of its
  energy eigenvalues (see Text). (b)
  Three-vertex hydra graph corresponding to
  the step potential above.

\bigskip\noindent 
{\bf Fig.~3:} The staircase function $N(k)$ and the average staircase
$\bar N(k)$. For the regular graphs the average staircase intersects every
``stair'' of the $N(k)$ graph, with the separators (\ref{sep}) showing as the
intersection points $N(\bar k_{n})=\bar N(\bar k_{n})$.

\bigskip\noindent 
{\bf Fig.~4:} Relative error $\epsilon_n^{(q)}=
|k_n^{(q)}-k_n^{({\rm exact})}|/k_n^{({\rm exact})}$ 
of (\ref{kn}) (see text) 
by including periodic orbits up to length $q$. 
The three curves shown correspond to 
$k_1$, $k_{10}$ and $k_{100}$ as indicated in the figure.

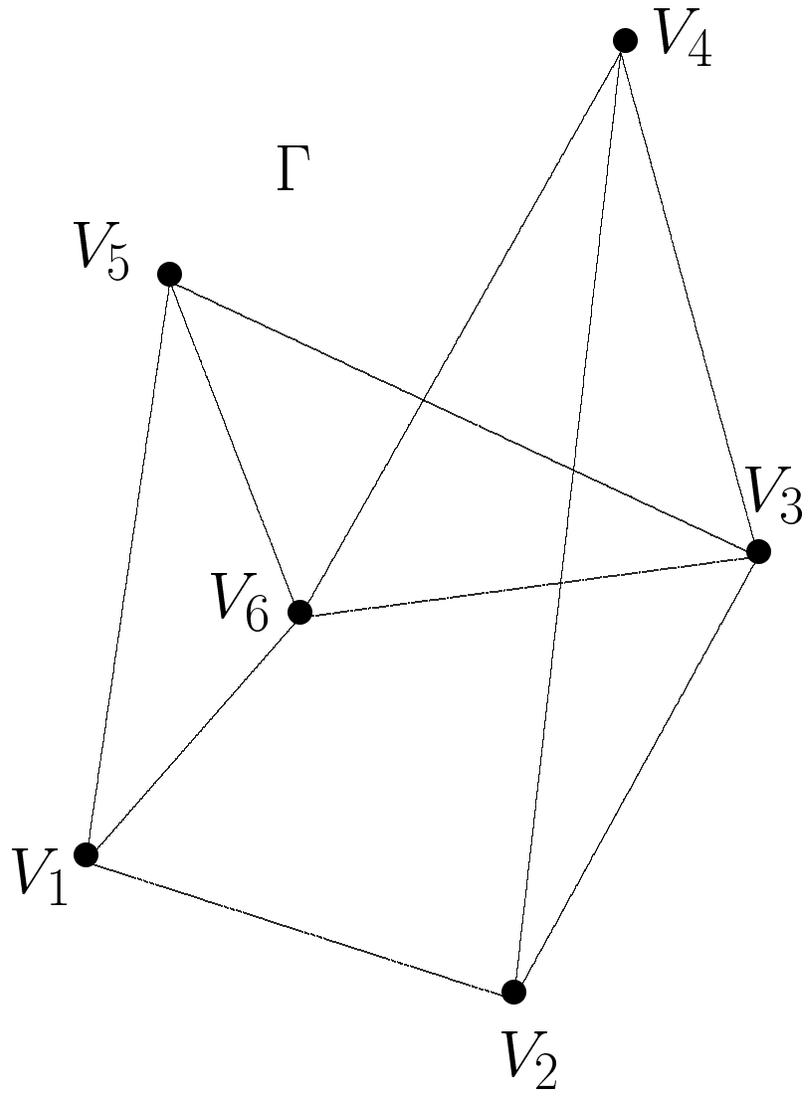
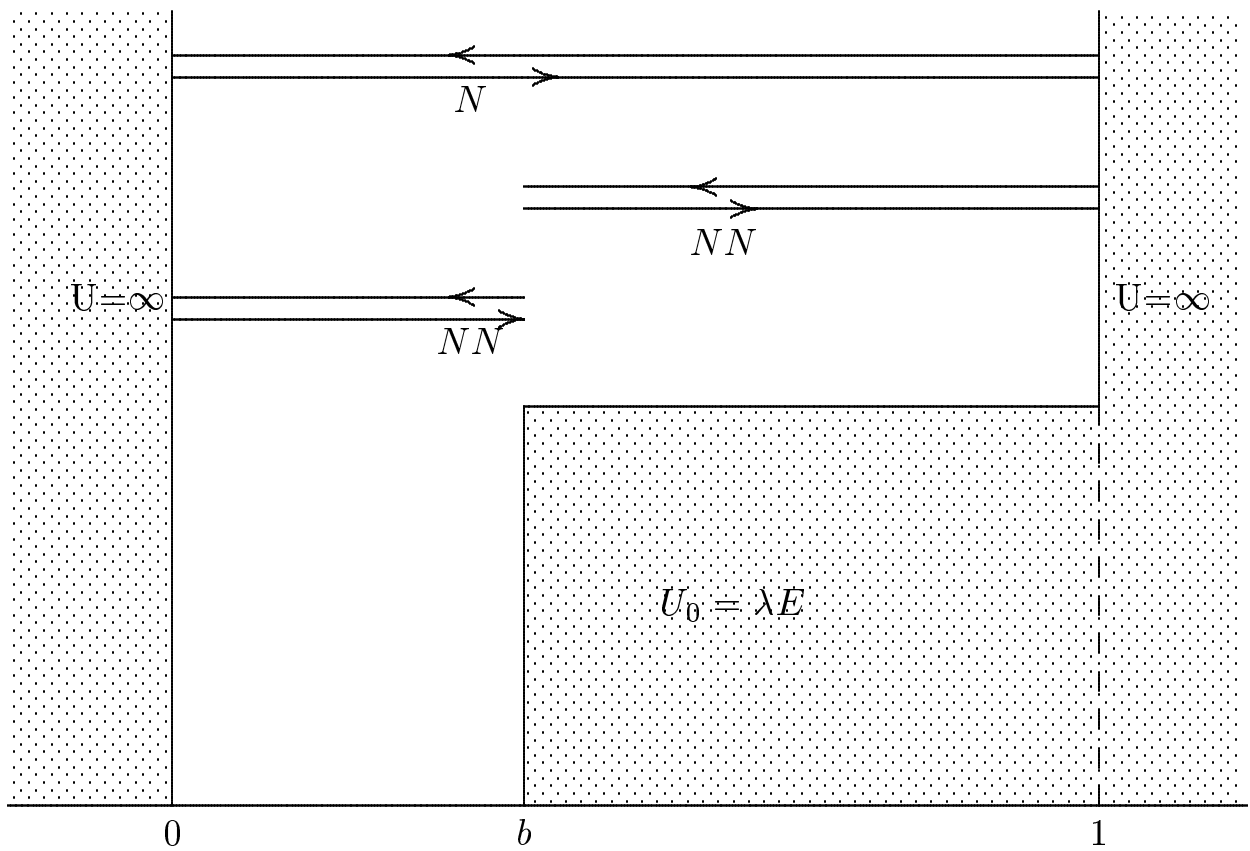
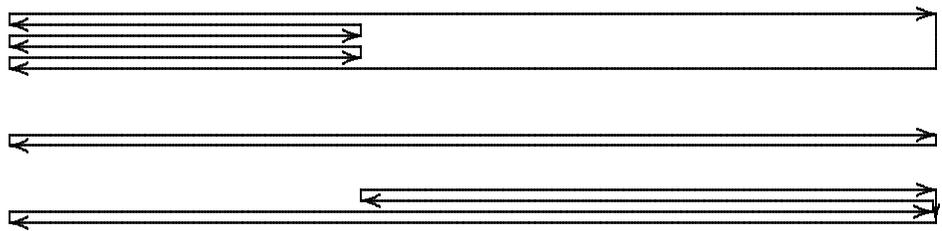
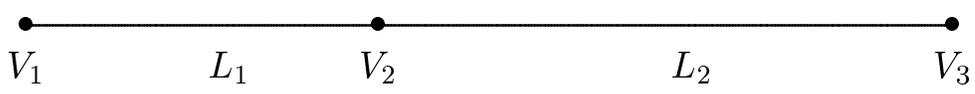
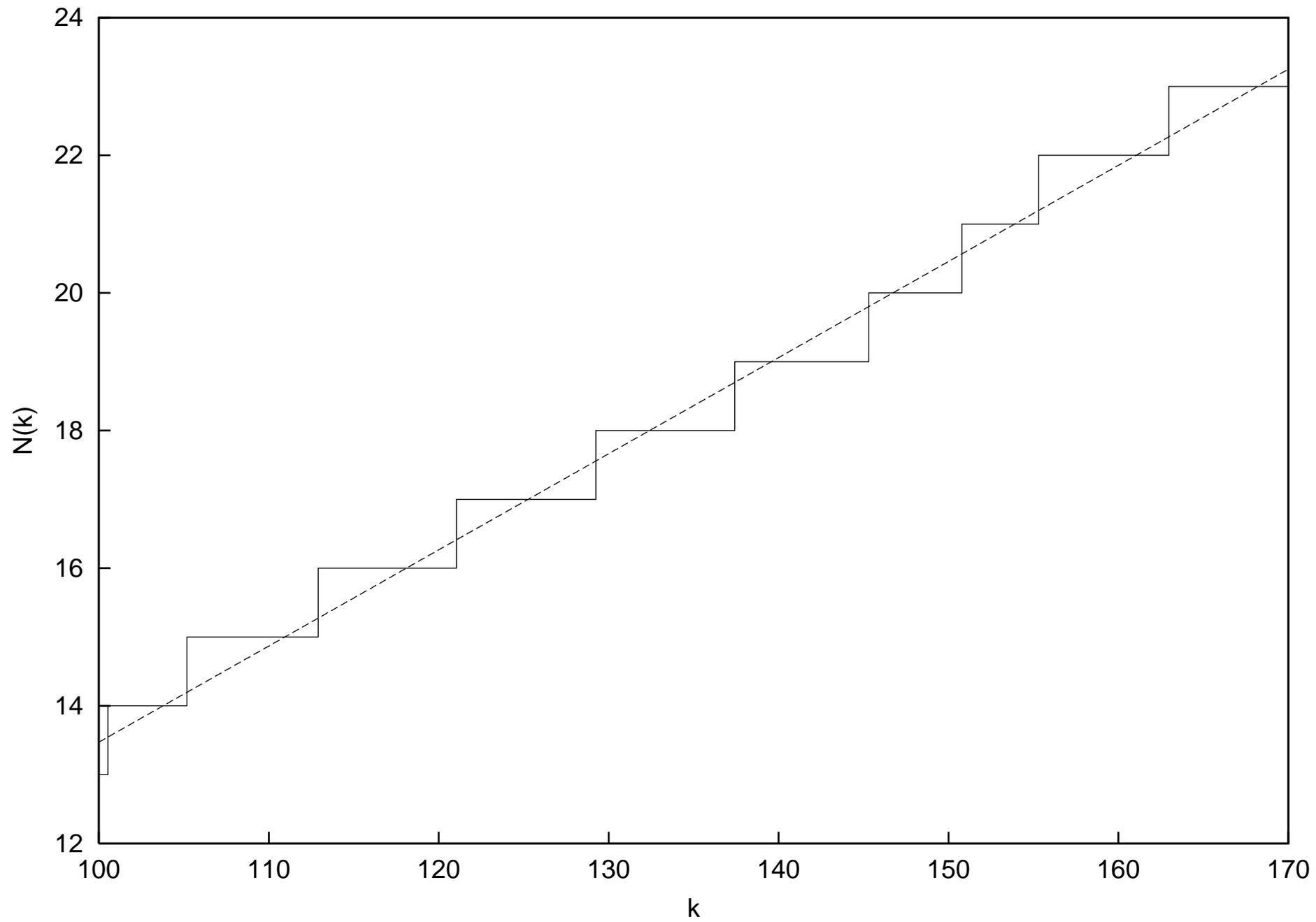
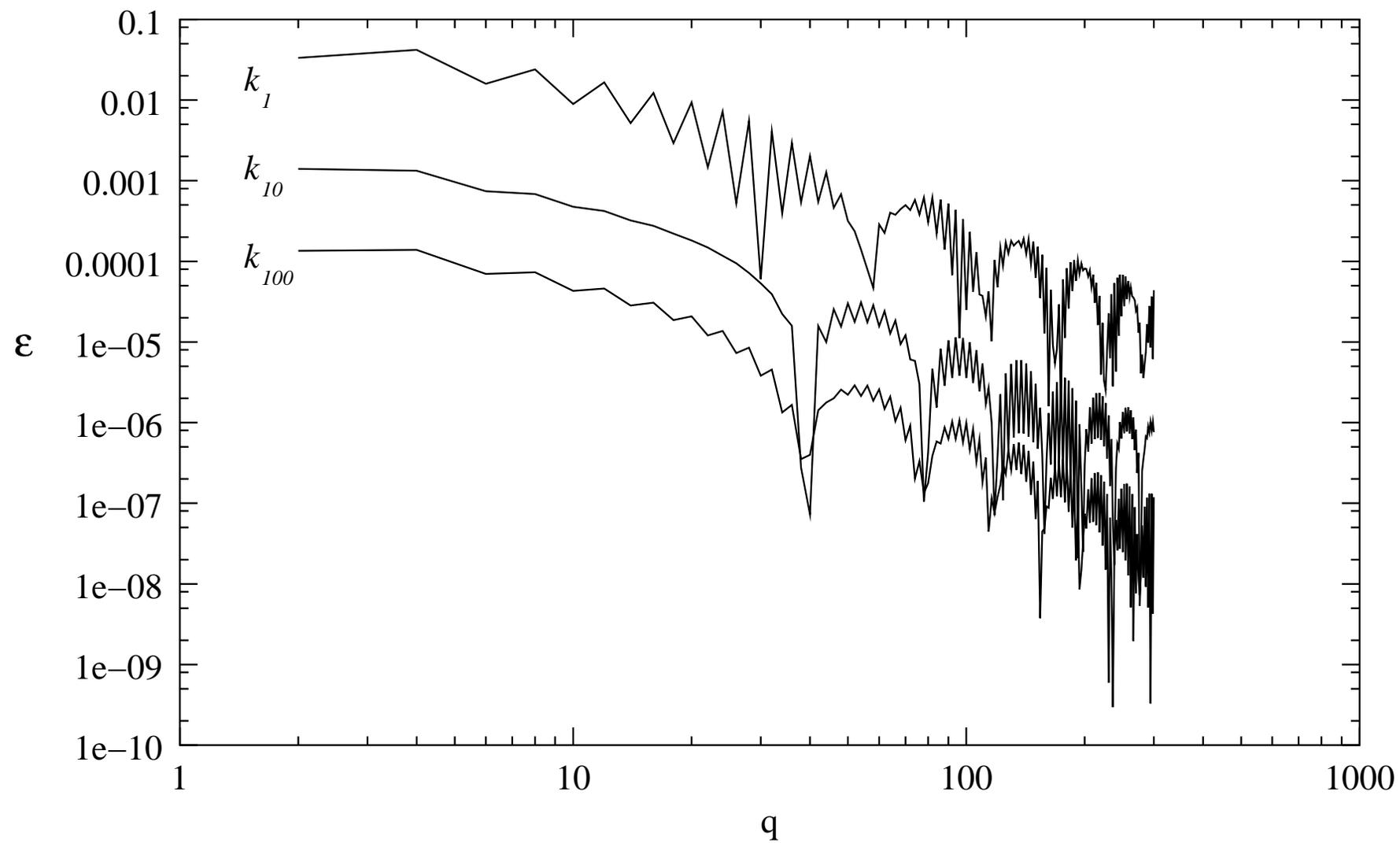
\end{document}